\documentstyle[psfig,proceedings]{crckapb} 
\def\l{$\lambda$}
\def \m{\ifmmode M_\odot\else M$_\odot$\fi}

\def\etal{{\it et al.}}
\def\ni{$^{56}$Ni}
\def\co{$^{56}$Co}
\def \lta {\mathrel{\vcenter
     {\hbox{$<$}\nointerlineskip\hbox{$\sim$}}}} % less than approx
\def \m{\ifmmode M_\odot\else M$_\odot$\fi}
\def \L {\ifmmode L_\odot\else L$_\odot$\fi}
\def \etal {{\it et~al.}} 
\def\deg{\ifmmode^\circ\else$^\circ$\fi} 
\def \Teff{{\it T}\lower.5ex\hbox{\rm eff}}

\begin{opening}
\title{NEW PERSPECTIVES ON TYPE IC SUPERNOVAE\protect\\
       }
\author{A. Clocchiatti}
\institute{Cerro Tololo Interamerican Observatory\\
          La Serena, Chile}
\author{J. C. Wheeler}
\institute{Department of Astronomy, University of Texas\\
          Austin, TX 78712}
\end{opening}

\runningtitle{TYPE IC SUPERNOVAE}

\begin{document}

% The \begin{document} command comes after the \end{opening}
% command.

\begin{abstract}

The Type Ic supernovae are probably ``nearly-nude" core collapse events.
They are thus of special interest in terms of their 
evolutionary origin, but also as the source of potential clues to
the explosion process that may not be obtained from other supernovae.
Study of Type Ic shows that they are distinct from Type Ia, even
those of the dimmer, redder variety.  
Type Ic also suggest that
the historical classification scheme based on spectra cannot encompass
the physical features of these and closely related events.  
The Type Ic
probably all have a very small outer layer of helium, but not enough to 
make them merely a version of Type Ib.  They probably require
another evolutionary channel.
Similar spectra near maximum light
belie a variety of late-time light curve behavior.  
Some events classified as Type Ic, eg SN 1962L, 1983V, 1990B, and 1990U, 
follow a relatively slow evolution
after maximum light, decaying $\sim$3.0 magnitudes in V during the first 150
days after maximum.  
These events seem to belong to a general category 
that includes SN 1993J and the Type Ib SN 19983N.  Their light
curves decline more rapidly than the expected late-time $^{56}$Co
decay with full trapping, but less rapidly than Type Ia.  
The late
time light curve of these events seems to be independent of the 
velocity of the ejecta and hence implies the trapping of $\gamma$-rays
in a manner that is independent of the mass that participates in
the homologous expansion. 
Other Type Ic, SNe 1983I, 1987M, and 1994I,
show steeper decline than even Type Ia.

\end{abstract}

\section{Introduction}
\label{se:cw:introduction}

Type~Ic supernovae (SN~Ic) have been 
spectroscopically defined by the absence of the strong H lines,
the strong SiII \l\l6347,6371 (hereafter SiII \l 6355) blend,
and the strong HeI lines, that characterize SNe of Type~II, Ia, and Ib,
respectively.
There are a variety of reasons to study SN~Ic.  
As has now become abundantly clear, Type Ia (SN~Ia) are not strictly
homogeneous in terms of their
light curve and spectral evolution behavior.  Some
SN~Ia have dimmer, redder light curves that are qualitatively similar
to a description of SN~Ic.  
The ejected mass in SN~Ic is probably low
and this could be connected with sub-Chandrasekhar mass ``edge-lit" models
of SN~Ia.  
To understand the progenitor evolution and the physics
of the explosion of both types of supernovae, one must understand
the differences as well as the similarities.  
It is also important
to distinguish SN~Ic from SN~Ia in the context of cosmological studies
that seek to use the latter as probes of the distance scale. 
A second reason to study SN~Ic supernovae is to understand their
progenitor evolution.  
We would like to know what distinguishes
them from Type II (SN~II), Type Ib (SN~Ib), and transition events
like SN 1993J, if, indeed, all of these are core-collapse events.
There is  a strong suspicion that SN~Ic must arise in binary systems,
but little understanding of how this happens. 
There is certainly no proof that they do so.  
An important reason to study SN~Ic 
is that they very probably represent ``nearly-nude" core collapse. 
They seem to be stripped of both their hydrogen and helium
layers, and to have rather small ejecta mass.  
They may bring us
closer than any other class of supernovae to directly witnessing 
the outcome of core collapse.  
The most tantalizing reason to study SN~Ic is that they may teach us
to look at all supernovae with new eyes.  
They seem to be presenting us with some surprises
and challenges to preconceived notions that may lead to deeper understanding.

The first step toward understanding in astronomy and many other 
sciences is to construct a taxonomy, a classification scheme that
allows us to give some semblance of order to nature as a guide to thinking
more deeply about it.  
The goal, of course, is an understanding
of the physics, not the construction of beautiful classification
systems.  
Astronomy is replete with the the leftovers of previous
fruitful attempts to classify that have led to understanding that
overthrew the original classification scheme.  
One of the more famous
is encapsulated in the Hertzsprung-Russell diagram.  This is
still a powerful tool for stellar evolution despite the fact that
the spectral types in a physically meaningful sequence 
so violate the alphabet that 
mnemonics are required to recall the order and the axis is forever
destined to run backwards in temperature.  
The SN~Ic may be pointing
the way to such a breakdown in our fifty year-old supernova classification
scheme as we strive the learn the lessons they teach.  

A certain humility about classification schemes is taught in a short story
by Jorge Luis Borges (1964), in which is described a Chinese encyclopedia
that divides animals into the following categories: 
(a) belonging to the Emperor,
(b) embalmed, (c) tame, (d) sucking pigs, (e) sirens, (f) fabulous,
(g) stray dogs, (h) included in the present classification,
(i) frenzied, (j) innumerable, (k) drawn with a very fine
camelhair brush, (l) {\it et cetera}, (m) having just broken
the water pitcher, (n) that from a long way off look like flies.
SN~Ic may be leading to a quixotic taxonomy, but one that may
ultimately bring deeper understanding.  
Perhaps the animals drawn
with a camelhair brush will cease to break the water pitcher.
We can hope, at least, to come to a scheme that is not solely 
included in category (h).

We give a brief comparison of SN~Ic and SN~Ia in \S~\ref{se:cw:SNIa}.
Section \ref{se:cw:current_status} gives a brief history of the literature
on the spectral spectral classification of SN~Ic.
In \S~\ref{se:cw:HHe} we discuss arguments about the existence of
H in SN~Ib as background for the discussion of SN~Ic.
Section \ref{se:cw:IcHe} presents arguments that all SN~Ic may, indeed, contain
small amounts of He.
In \S~\ref{se:cw:light_curves} we present the 
collected photometric and spectroscopic
results on several well-observed SN~Ic.
In \S~\ref{se:cw:tax} we describe the
combined light curves and outline the arguments for the
need of a new taxonomy for SN Ic and related events.
In \S~\ref{se:cw:disc} we discuss the results and summarize our conclusions.

\section{Type Ia Supernovae}
\label{se:cw:SNIa}

This review will concentrate on SN~Ic, so we will make only a few
comments on SN~Ia in this context.  
The dimmer, redder
SN~Ia show some spectral similarities to SN~Ic, but most of this
is incidental and due to the fact that cool photospheres tend
to show ubiquitous lines of Fe II which are not, at the current
level of sophistication, useful for diagnostic purposes.  
Closer examination
shows that the spectra of SN~Ic can be distinguished from SN~Ia,
even near maximum light.  There is little similarity at later times.

Although the edge-lit models
for SN~Ia (Woosley and Weaver 1994; Woosley this volume)
are plausible in terms of their evolution and physics,
nature apparently does not produce them in the observed sample of SN~Ia.  
They seem, in particular, to be too blue compared to the
observed events, especially the dim ones (H\"oflich \etal, this volume;
H\"oflich and Khokhlov 1996; Wheeler 1996).  
It seems that dim, red SN~Ia events can be reproduced
rather well both in terms of their multi-color light curves and
maximum light spectra by models of the
``pulsating delayed detonation" variety that are based on carbon/oxygen
white dwarfs of the Chandrasekhar mass that eject a rather small
mass of  $^{56}$Ni (H\"oflich \etal, this volume;
H\"oflich, Khokhlov and Wheeler 1995; Wheeler \etal\ 1995).

There is a clear area in which the Chandrasekhar mass models
for SN~Ia require some more thought, the slope of the late-time
tail.  There is no question that current models cannot easily
account for the extensive slope of about 56 day half-life
observed in several SN~Ia, as emphasized by the excellent
review by Colgate, Fryer and Hand in this volume.  One way
to characterize this problem is to look at the ratio of
luminosities at the peak and on the tail.  This ratio
is distance independent.  Under the assumption  
that all $\gamma$-rays are trapped at maximum and all
positrons are trapped on the tail, there is a lower limit
to the ratio of the luminosity on the tail to that at
maximum that is also independent of the \ni\ mass. This limit
depends somewhat on the time of maximum.  For a maximum at
20 days it is 
log L$_{max}$/L$_{tail}$ $\leq$ 1.53 + t(days)/256.  Several 
observed SN~Ia show a peak/tail contrast in excess of
this amount.  It is first violated, at least in V,
in SN 1992A at about 150 days and by 600 days the observed
light curve is about a factor of 5 dimmer than that
predicted by full trapping of positrons (Kirshner \etal\ 1996).
Either this light is escaping in an undetected band
or not all the positron energy is trapped.  
Colgate \etal\ are correct in stressing the reality
and importance of this problem. Their solution,
involving the collapse of a core and ejection of small
nickel and ejecta mass, is, however, unlikely to be correct.
The peak is probably not as insensitive to 
ejecta mass as their one-zone models suggest.  
Their small ejecta mass collapse models may be a quite
reasonable representation of the SN~Ic events with
``fast" light curves, as described below.

\section{Type~Ic Supernovae and Related Events}
\label{se:cw:current_status}

Type Ic supernovae were proposed as a taxonomical class by 
Harkness \& Wheeler (1990) and Wheeler \& Harkness (1990)
to codify the original categories ``helium-rich'' and 
``helium-poor'' (Wheeler \etal\ 1987) for SN~Ib.
The reality of SN~Ic as a distinct subtype has been constantly challenged, on
the grounds that there could be a gradient of He abundances or
excitation in SN~Ib
which could include SN~Ic as a one manifestation
(e.g. Branch, Nomoto \& Filippenko 1991; Leibundgut 1995).
The original perception of SN~Ib as a separate group was
based on spectroscopic observations (Wheeler \& Levreault 1985),
light curves in the infrared (Elias \etal\ 1985), and absolute
brightness at maximum (Uomoto \& Kirshner 1985, and references therein).
All of these results were published nearly simultaneously, and their combined
weight was strong enough to establish beyond any doubt
that SN~Ib were a group separate from those of SN~Ia
on probable physical as well as clear taxonomic grounds.
In contrast, the differentiation of SN~Ic as a separate group
has been based on spectroscopic properties alone
(Wheeler \etal\ 1994, and references therein),
while the properties of the light curves were almost completely ignored.
This limited basis, together with the similarity of spectra 
of SN~Ib and SN~Ic at late times and the supposed
uniformity of the light curves in the infrared (Elias \etal\ 1985, included
two spectroscopic SN~Ib and one SN~Ic in their infrared light curves),
have all contributed to make the reality of SN~Ic difficult to establish
beyond reasonable doubt.
Spectroscopic analyses of SN~1987M
(Jeffery \etal\ 1991; Swartz \etal\ 1993a; Wheeler at al. 1994) 
reach different conclusions based on essentially the same explosion
model and comparison with the same set of observed spectra (Filippenko, Porter,
\& Sargent 1990).
On the other hand, early and late--time light curves of SNe prompt different
kinds of models in the few cases in which such a study has been attempted
(Ensman \& Woosley 1988; Swartz \& Wheeler 1991).

It is generally accepted now that NLTE effects are crucial
to populate the excited He levels and that fast electrons originating
in the scattering of $\gamma$-rays from the decay of $^{56}$Ni and $^{56}$Co
are an obvious source of those electrons (Harkness \etal\ 1987; Lucy 1991;
Swartz \etal\ 1993a).
Accordingly, the detection of He lines implies that the He layers are exposed to the
$^{56}$Ni and, probably, some sort of large scale dynamical instability which
either mixes the $^{56}$Ni outwards or breaks down the intervening mass
layers, opening windows between the He and Ni regions.
Since the strength of the He lines will result from the combination of
abundance and NLTE excitation, the degree of mixing required by a theoretical
spectrum to match the observations is strongly dependent on the model chosen,
and spectral modeling cannot provide hard limits on the degree of
mixing or He abundance.

Although the spectroscopic detection of He lines alone
could require some degree of mixing, most of
the suggestions for extensive mixing of the $^{56}$Ni outwards 
in SN~Ib/c have originated
from theoretical studies of the light curves.
A decreased escape probability for $\gamma$-rays was suggested by 
Ensman \& Woosley (1988) in order to reproduce 
the light curve of SN~1983N around
maximum light while mixing was invoked by Shigeyama \etal\ (1990)
to explain the light curves of SNe~1983N and 1983I.
Both analytical and numerical studies of the mixing led 
Hachisu \etal\ (1991; see also Nomoto, Filippenko \&
Shigeyama 1990; Jeffery \etal\ 1991) to associate
low mass He stars which undergo extensive mixing with the fast photometric
evolution of SN~Ic, while objects with more slowly evolving light
curves like SNe~1983N and 1984L would have experienced less mixing.
This scenario, however, had the drawback that the more massive objects would
accordingly show less NLTE effects and weaker He lines, and the less massive
objects would show more NLTE effects and stronger He lines,
in contradiction with the observations of SN~1984L. 
SN~1984L
showed strong He lines at early times (Harkness \etal\ 1987; Lucy 1991) and a
very slow exponential decay (Schlegel \& Kirshner 1989) consistent with an
envelope as massive as 10 \m (Swartz \& Wheeler 1991).
Baron (1992) emphasized this contradiction and suggested
that the identification of the models should be reversed.
In his view,
both SN Ib and Ic would come from He stars exploding after evolving in
interacting binary systems, but the former have to be associated with the
low mass and the latter with the high mass He star models.
Baron proposed that the higher photospheric velocities of the
hypothesized more massive SN~Ic
progenitors compared to the less massive SN~Ib would be the
result of higher explosion energies in SN~Ic.

The He star model
has been partially abandoned after the popularization of C$+$O stars as
a suitable model for SN~Ic (Wheeler, Swartz, \& Harkness 1993;
Nomoto \etal\ 1994, Iwamoto \etal\ 1994).
These models lack the density contrast of the He/C$+$O interface, and large
scale mixing due to Rayleigh--Taylor instability is not expected to be
important.
In addition, the low mass of the ejecta and its very fast expansion velocity
would dominate the decrease of the $\gamma$--ray deposition function and
would make the effect of microscopic  mixing of \ni\ nearly irrelevant.

The contradiction between light curve and spectral modeling surfaced again
in the case of SN~1993J, a SN~II SN closely related to SN~Ib/c.
Detailed study of the spectrum at the onset of the nebular stage
(Swartz \etal\ 1993b)
suggested that mixing was not needed, and could not have been extensive
or else H and He lines would be much stronger than observed.
On the other hand,
detailed study of the bolometric and monochromatic light curves
(Woosley \etal\ 1994) prompts the necessity of a
moderate degree of mixing or else the light curve will show sharp edges
as the photosphere evolves through regions of low optical depth
(see also discussion in Wheeler and Filippenko 1996; H\"oflich \etal\
1996).

There has been an accumulation of observational
results and associated theoretical studies which allows a new view of the
issues surrounding SN~Ib/c.
On the theoretical side, model atmospheres of low mass He stars (Swartz \etal\
1993a; Wheeler, Swartz, \& Harkness 1993) 
produced convincing evidence that these
stars could not match the spectra of SN~Ic, either at maximum or during
the early nebular phase, and that nearly pure oxygen atmospheres gave better
agreement with the observations.
An effort has also been made to compute the
outcome of stellar evolution both for massive single stars and stars in
interacting binary systems, in order to have a better idea of what sort of
progenitor
could be expected from these scenarios (Woosley \etal\ 1994; Woosley, Langer,
\& Weaver 1995).
A key issue remains the amount of helium that could be allowed
in the outer layers of SN~Ic and thus whether they are a natural 
extension of SN~Ib to envelopes of low helium mass as might result
from extensive Case B mass exchange (Woosley, Langer,
\& Weaver 1995; Woosley, these proceedings) 
or whether they require a more severe loss of
helium envelope mass as might result from common envelope 
and ``Case BB" evolution in which the helium envelope also
swells to fill its Roche lobe and transfer mass to a companion
(Nomoto \etal\ 1994, Iwamoto \etal\ 1994; Nomoto, these proceedings).
The former cannot easily provide a helium layer less than 0.1\m,
whereas the latter has some promise for doing so.

On the observational side, there have been well-observed
new events and reduction
and study of observations of important old events which remained unexplored for
more than a decade.
Astronomers at Asiago
Astrophysical Observatory and ESO provided detailed late--time photometry of
SNe 1990B and 1990U, two moderately bright events with a SN~Ic spectrum
(Benetti \etal\ 1995).
SN~1993J provided the equivalent of a 
``Rosetta Stone'' in supernova spectroscopy
by undergoing a well-documented transition from a SN~II to a hybrid between a
SN~II and SN~Ib spectrum
(Filippenko \& Matheson 1993; Wheeler \& Filippenko 1996, and references
therein).  Then SN~1994I brought the outstanding opportunity
to observe a bright SN~Ic beginning one week before maximum light and
provided keys to understanding the evolution of He lines in SN~Ic events
(Filippenko \etal\ 1995; Clocchiatti \etal\ 1996a).

Supplementing these recent observations, 
late--time CCD photometry of SN~Ib 1983N has finally
settled the open question of its exponential decay (Clocchiatti \etal\ 1996b),
and an extensive set of
observations of the bright SN~Ic SN~1983V has finally been presented
(Clocchiatti \etal\ 1996c).
The combined data of all these objects allows two distinctly striking
conclusions: a) Irrespective of very similar spectroscopic evolution,
not all SN~Ic have similar photometric evolution,
and b) irrespective of the very different spectra near maximum light, 
some SN~Ib, SN~Ic, and hybrid SN~II (called IIb in some
quarters) have very similar photometric evolution.

\section{The Possible Presence of H in SNIb/c}
\label{se:cw:HHe}

Near its second maximum, SN~1993J showed a small absorption
of He I \l6678 on top of the emission component of a strong
P-Cygni profile of H$\alpha$. 
 Wheeler \etal\ (1994) noted that
a qualitatively similar structure existed in the SN~Ib 1983N and 1984L
and perhaps in the SN~Ic 1987M and 1994I, with increasingly high
velocities as one went from SN~1993J to SN~Ib to SN~Ic.  
They studied the photospheric spectra of these events with simple
power-law atmospheres.  
Their conclusion was that the SN~Ib may,
indeed, contain a small outer layer of H, perhaps 0.001 \m, at
high velocity, but that the corresponding features in the SN~Ic were
due to SiII~\l6355 and probably CII~\l6580.  They noted that
the flat bottom of the H$\alpha$ absorption in SN~1993J just before
the helium appears may also be due to SiII~\l6355.

For SNe~1987M and 1994I Wheeler \etal\
used model atmospheres based on an expanding layer of oxygen with solar
abundances of heavier elements and an outer shell of He with solar abundance
of heavier elements.
They analyzed, in particular, the possibility that the absorption line near
6370\AA\ in the spectrum of SN~1987M seven days past maximum light
were HeI~\l6678 {\em and} that the Na~D line blends with HeI~\l5876.
They concluded that two minima should appear near the position of Na~D, where
SN~1987M shows only one and, hence, that the optical 
photospheric spectra of both
SNe~1987M and 1994I were consistent with no He.
Discovery of He in SN~1994I required a revision of this conclusion.

\section{Helium in SN Ic}
\label{se:cw:IcHe}

\begin{figure}
\psfig{figure=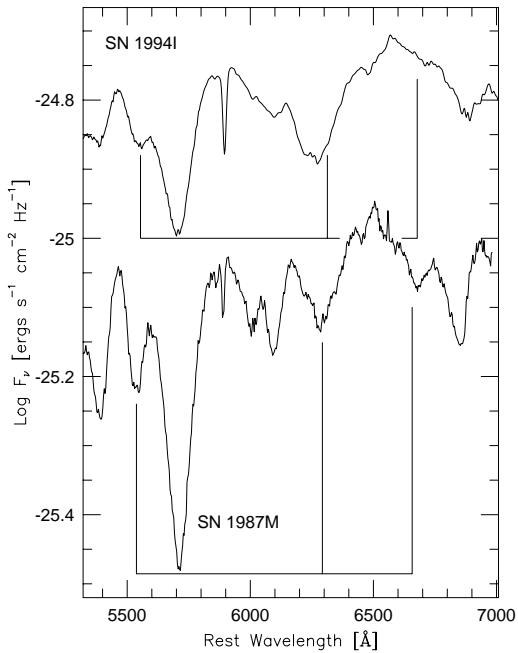,height=4in}
\caption{\label{fi:94iHe} The HeI optical series in SN~1994I and SN~1987M.
The redshift of the parent galaxies have been removed.
The thick solid lines mark the position where the lines of the HeI optical
series should appear, blueshifted by 16,900 km~s$^{-1}$ in SN~1994I and
17800 km~s$^{-1}$ in SN~1987M.}
\end{figure}

Filippenko \etal\ (1995) show that SN~1994I displayed a strong HeI~\l10830
line, as had been predicted by Jeffery \etal\ (1991), Swartz \etal\ (1993a),
and Wheeler \etal\ (1994) using different models intended to represent a
SN~Ic.
The expansion velocity of HeI forming this IR line perfectly 
matches the velocity of the HeI needed to form the small and 
transient absorption feature that Clocchiatti \etal\ (1996a)
identify as HeI~\l5876.
By comparing the spectra of SNe~1994I and 1987M, Clocchiatti \etal\
were also able to identify a
much stronger feature in the latter as a candidate to be HeI~\l5876, and then
two other lines of the HeI optical series at nearly the same expansion velocity.
This is illustrated in Figure \ref{fi:94iHe}.

%%%%%%%%%%%%% Figure 1 here %%%%%%%%%%%%
%% Fig 3 of 94I Helium Paper

HeI~\l5876 in SN~1994I carves the blue
side of the Na~D absorption and shifts the centroid of the resulting blend
to the blue.  This behavior gives strength to the interpretation 
of Clocchiatti \etal\ (1996c) who
take this effect as evidence for He in SN~1983V.
In addition, by inspecting Fig.~5 in Filippenko (1988) it is possible to
see that SN~1988L also displayed the HeI line series on 1988~May 11.
HeI~\l 5876 appears at $\sim 5550$\AA\ (-17,100 km~s$^{-1}$).
HeI~\l 6678, HeI~\l7065, and HeI~\l7281 are also seen at approximately
the same expansion velocity.
The minimum of HeI~\l7065, however, is almost completely filled by
[SII]~\l\l6713,6734 emitted in a nearby HII region.

Swartz \etal\ concluded that the total mass in He and H 
in the outer layers of SN~1987M should be smaller than 0.1~\m.
From the velocities of helium measured by Filippenko \etal\ (1995)
and Clocchiatti \etal\ (1996a) one can estimate a characteristic
energy.  
The velocity is about 17,000~km s$^{-1}$.  
To move 0.1 \m\ of helium at that velocity requires an energy of 
$\sim3\times10^{50}$ ergs, a substantial portion of the total kinetic energy
expected from a supernova.  This gives another strong indication
that the mass of helium ejected in the outer layers is
substantially less than 0.1~\m.

The discovery of HeI lines in the SNe 1983V, 1987M, 1988L and 1994I 
terminates the controversy on whether some 
He is physically present in SN~Ic events.
The next question is the relevance of that helium.
The presently preferred model 
for SN~Ic of a C$+$O star which evolved as an exposed core
through mass loss to an interacting binary companion and/or strong winds
is consistent with the presence of some He in the external layers
(Iwamoto \etal\ 1994; Woosley, Langer \& Weaver 1995).
The {\em amount} and {\em dynamics} of that
helium, may however, be an important discriminant among different
evolutionary processes.
From what little is known, one is led to suggest that interacting binary
evolution would produce a bi-modal (or even multi-modal) distribution
of helium envelope mass
rather than a continuum of progenitors with smooth
variation of physical properties.  It is likely that SN~1984L has
a large helium envelope, SN~1983N a rather modest one, and
SN~1994I one that is distinctly smaller yet.

It seems unlikely that one can avoid this
conclusion with careful manipulation of the NLTE excitation of
a larger mass of helium.  
The helium that is likely to be excited is that
closer to the source of excitation in the \ni\ or \co\ and that will
be in lower layers at lower velocities than the observed 17,000 km s$^{-1}$.
In addition, any mass of helium will become more exposed to the radioactive
excitation at later times.  
Any model of SN~Ic with a mass of helium
in excess of 0.1 \m\ must pass the test of SN~1987M that no lines of
helium are observed about 60 days after maximum when Swartz \etal\ argue
they would be at their maximum strength in a helium star model.

These arguments lead us to conclude that although there is probably some
small mass of helium, $\lta$ 0.01 \m, in the outer layers of all SN~Ic,
this is substantially less than must be present in SN~Ib like
SN~1983N and especially SN~1984L.  
The most plausible hypothesis
still seems to be that SN~Ic require a substantially different
evolution than SN~Ib and that something like Case BB mass transfer,
rather than Case B mass transfer, is required to understand them.
There is probably hydrogen in SN~Ib, but that does not make them
a SN~II.  There is helium in some, probably all, SN~Ic, but that
does not make them a SN~Ib.  This points to a different progenitor
evolution for SN~Ib and SN~Ic.  On the other hand, spectral evolution
does not tell the whole story.  Some SN~Ib and SN~Ic share light
curve behavior and hence some common physical basis.

\section{Light Curves of SN Ic and Related Objects}
\label{se:cw:light_curves}

\begin{figure}
\psfig{figure=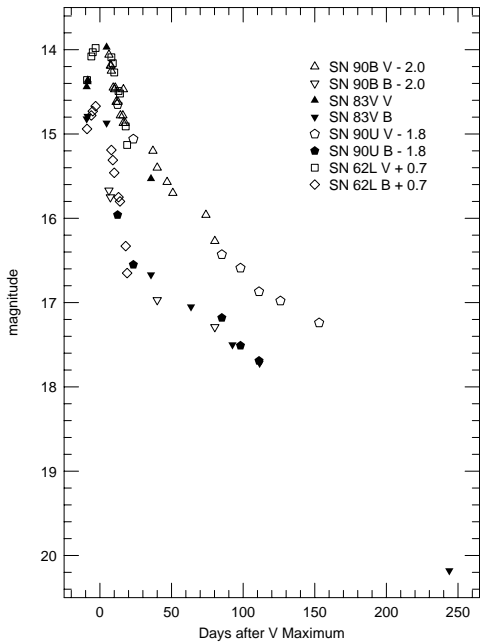,height=4in}
\caption{\label{fi:lcs} Light curves of SNe~1962L, 1983V, 1990B, and 1990U.
The key of symbols and the off--set in magnitudes for each object 
with respect to SN 1983V are given on the
upper right hand corner of the figure.} 
\end{figure}

We present here a collection of observations
from the literature and data still unpublished.
In Figure~\ref{fi:lcs} we have plotted the
photoelectric and CCD photometry of SNe 1962L
(Bertola 1964; Bertola, Mammano, \& Perinotto 1965), 1983V
(Clocchiatti \etal\ 1996c), 1990B and 1990U
(Benetti \etal\ 1996).

%%%%%%%%%% Figure 2 here %%%%%%%%%%%%%
%%% Fig 1 of Comparison Paper

%%%%%%% Figure 3 here %%%%%%%%%%%%%%%%%
%%% Figure 3 of Comparison Paper

\begin{figure}
\psfig{figure=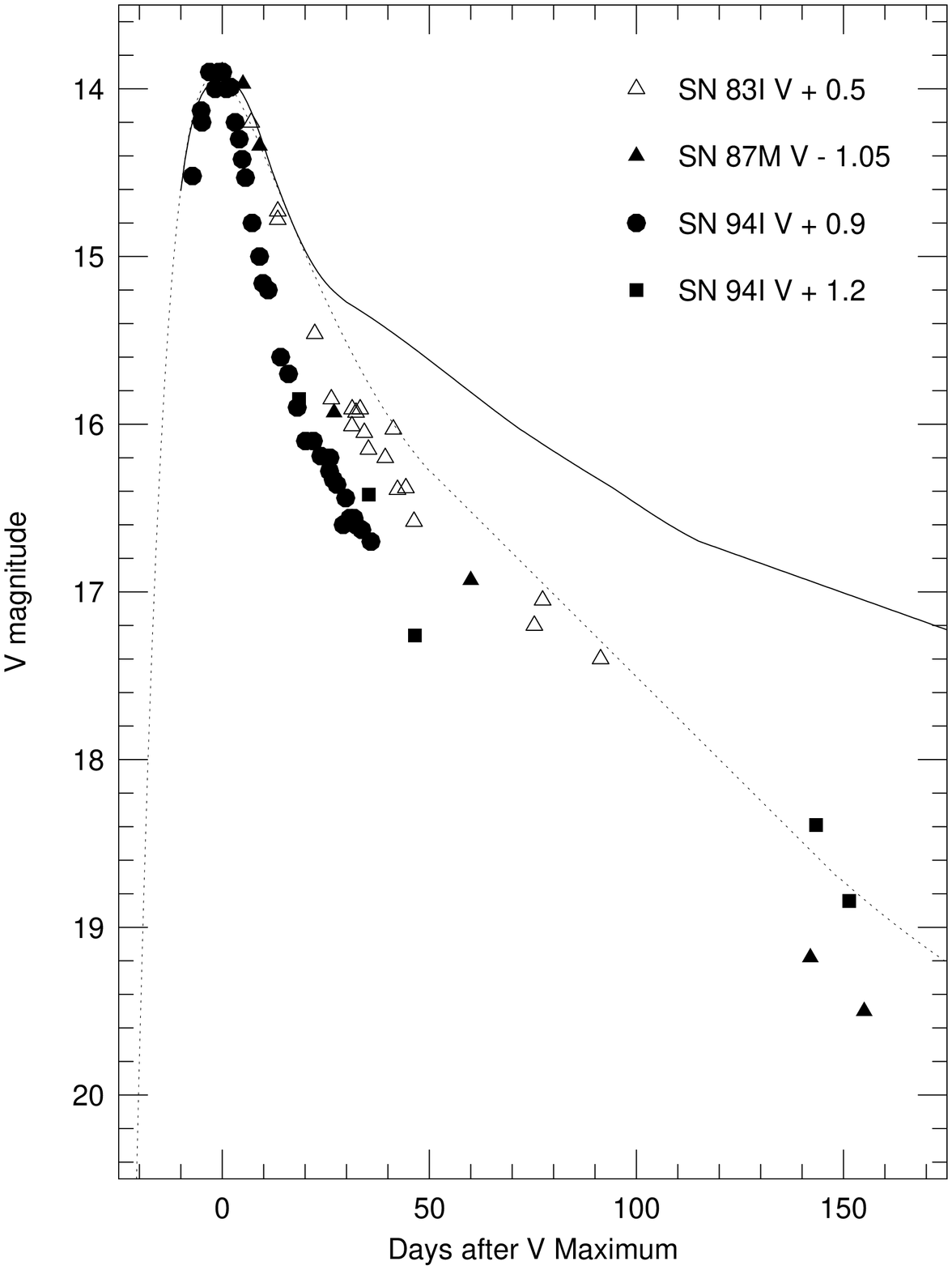,height=4in} 
\caption{\label{fi:lcf} V light curves of SNe~1983I, 1987M, and 1994I
compared with the template for the {\em slow} SN~Ic, 
and a template for SN~Ia (Wheeler \& Benetti 1996).} 
\end{figure}

In Figure~\ref{fi:lcf}, we show the V photometry of SNe~1983I and 1987M
together with data of SN~1994I distributed by the Center for Astrophysics
of Harvard University
(CFA; Schmidt \& Kirshner 1994), and data available in the archive of the
Royal Greenwich Observatory (RGO; Martin 1993).

Inspection and comparison of Figures~\ref{fi:lcs} and \ref{fi:lcf} reveal
a striking result.
The light curves of the SN~Ic plotted in Figure~\ref{fi:lcs} are
different from those plotted in Figure~\ref{fi:lcf}.
The former show a slower evolution than the latter.
After a maximum with FWHM on the order of 20 days, 
the SN~Ic in Figure~\ref{fi:lcs}
decay $\sim$1.2 magnitudes in V and $\sim$1.9 in B, 
and settle on the exponential
tail some 30 days after maximum.
The rates of decay on the tail are $\sim$0.018 mag day$^{-1}$ 
in V and $\sim$0.013
mag day$^{-1}$ in B, computed between 30 and 110 days after maximum.
The objects in Figure~\ref{fi:lcf}, by contrast, decay $\sim$2.0 magnitudes in V and
$\sim$2.6 or more in B before entering the exponential decay, also some 30 days
after maximum.
The few points of SN~1983I in V and SN~1987M in B and V
indicate that the decay of these objects in the tail was faster than that of
the objects in Figure~\ref{fi:lcs}.
The approximate rates of decay are 0.023 -- 0.029 mag day$^{-1}$ in V 
and is $\sim$0.021 mag day$^{-1}$ in B.

The evolution of the four objects in Figure~\ref{fi:lcs} is so
similar in the parts in which they overlap that the suggestion of a common 
locus of evolution immediately follows.
We have prepared a ``template'' light curve to represent this standard
evolution and the results are plotted in the figure.
The points beyond $\sim$110 days after maximum are uncertain, since they depend
only on one observation of SN~1983V in B and the late--time light curve of
SN~1990U in V.

The objects in Figure~\ref{fi:lcf}, on the other hand, do not provide a
clear basis for tracing a template.
By normalizing the maxima, as we have done in Figure~\ref{fi:lcf}, it
is not possible to adjust SN~1983I, 1987M and 1994I 
to the same locus of evolution.
This could be an indication that the objects in the fast SN~Ic subclass do not
constitute a homogeneous group, as the objects in the first one seem to do.

From the perspective of optical spectroscopy, SNe~1983V, 1987M, and 1994I
were extremely similar.
Indeed, any one of them could be taken as a perfect example of
what Wheeler \& Harkness (1990) and Harkness \& Wheeler (1990) meant by
a SN~Ic.
Careful comparison of the spectra reveals small differences.
Both fast SN~Ic 1987M and 1994I displayed a transient and weak HeI~\l5876 line.
On the other hand, the slow SN~Ic
1983V did not display HeI~\l5876 as a separate line,
and only the shift of the centroid of the blend NaI~D--HeI~\l5876 suggests that
the HeI line appeared between 18 and 38 days after maximum.
The fact that the fast SN~Ic seem to show stronger HeI lines may be a
direct consequence of their rapid evolution.
If the rapid decrease in brightness after maximum is a result of an enhanced
leaking of $\gamma$--rays, the small outer He layers will receive the required
non--thermal excitation before the fast expansion makes the He optical depth
too low to leave an imprint in the spectrum.
Slow SN~Ic might have less non--thermal excitation when the He
optical depth is still significant and miss the most favorable time to
produce He lines.

\section{A New Taxonomy?}
\label{se:cw:tax}

The similarity of the spectra of the SN~Ic near maximum
and the differences in the light curves
of Figures~\ref{fi:lcs} and ~\ref{fi:lcf} 
compose a remarkable contrast, and imply an interesting
counterexample to the way in which SNe have always been classified.
Since the very early stages of systematic research on SNe, more than
50 years ago, it has been stressed that the classification of SNe has to be
based on spectra and {\em not} on light curves.
The results described in the previous sections, however, prove that it would be
difficult to differentiate a fast SN~Ic
such as SN~1994I from a slow one like SN~1983V by analysis of a few spectra
taken near maximum light.

%%%%%%  Figure 4 here %%%%%%%%%%
%% Figure 1 of 83N 

\begin{figure}
\psfig{figure=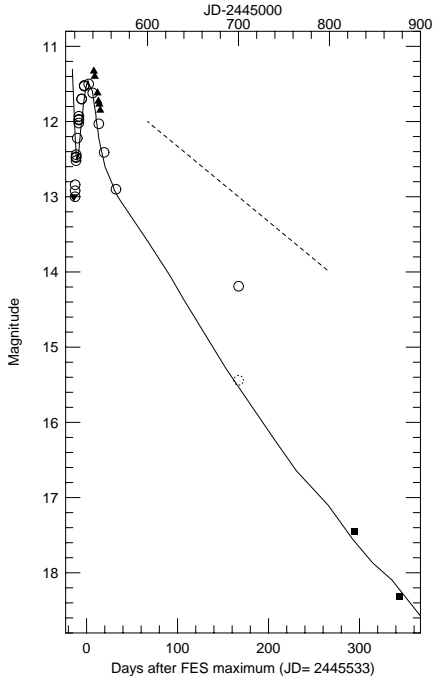,height=4in} 
\caption{\label{fi:83n93j}
The light curve of SN~1983N in the FES and V bands.
Solid open circles are FES
magnitudes and upright solid triangles are the photoelectric V points 
given by Panagia \etal\ (1985).
The dotted open circle is the late time FES point corrected
for the
brightness of the background according to Panagia (1994)
The downwards solid triangle is a, presumably
photoelectric, V magnitude (Thompson 1983).
The two solid squares are the CCD V magnitudes presented by
Clocchiatti \etal\ (1996b).
The solid line is the V light curve of SN~1993J obtained from the RGO database
(Martin 1993) to which we added 0.6 mag.
The dashed line is the slope expected for complete trapping of $\gamma$-rays
($\sim$ 0.01 mag day$^{-1}$).}
\end{figure}

%%%%%%  Figure 5 here %%%%%%%%%%
%% Figure 9 of 83V

\begin{figure}
\psfig{figure=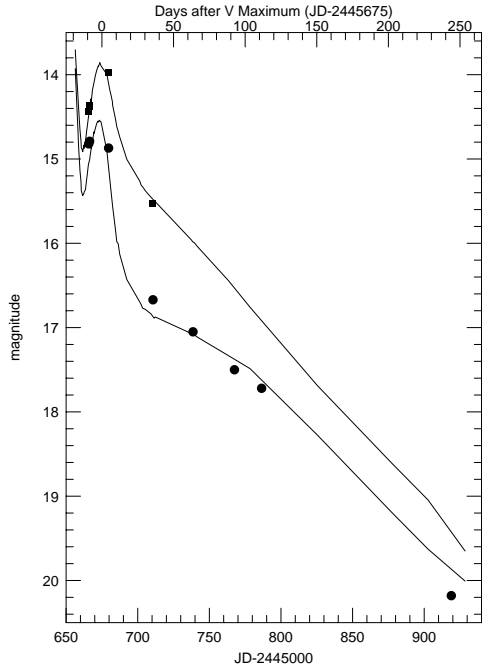,height=4in} 
\caption{\label{fi:83v93j}
Comparison of the light curves of SN~1983V (points) and
SN~1993J (solid line; Martin 1993). Squares are V and circles B.
The magnitudes of SN~1993J were
shifted by adding 3 magnitudes in V and 3.15 in B to compensate by differences
in brightness and color excess.}      
\end{figure}

%%%%%%  Figure 6 here %%%%%%%%%%
%% Figure 7 of Comparison

\begin{figure}
\psfig{figure=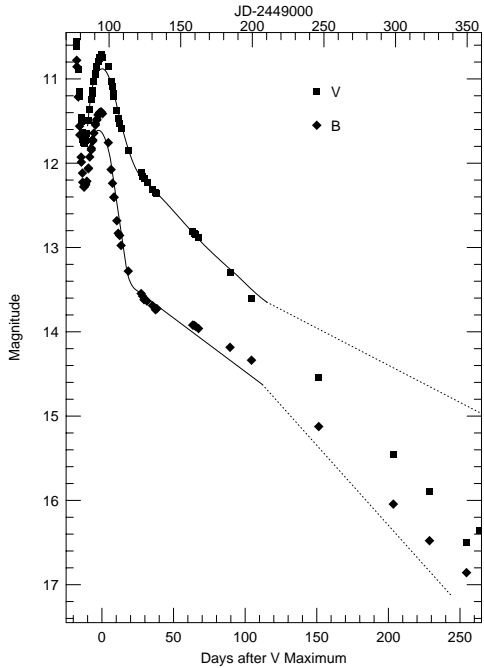,height=4in} 
\caption{\label{fi:93jyics} Comparison of the photometry of SN~1993J 
(Martin 1993), with the template light curves for the slow SN~Ic.
The templates beyond 115 days after maximum are plotted with dotted lines to
emphasize that they are more uncertain at these later phases, since they
rely only on SN~1983V in B and SN~1990U in V.
The Julian Day scale in the upper axis applies to SN~1993J.
The magnitudes of SN~1993J were
shifted by adding 3 magnitudes in V and 3.15 in B to compensate differences
in brightness and color excess.}     
\end{figure}

The separation in subclasses, however, is completely clear from the
photometric point of view.
In addition, slow
SN~Ic share characteristics with the SN~Ib
SN~1983N and the hybrid SN~1993J.
These two objects have, in turn, a similar evolution of their V light curves and
velocities for some 300 days after maximum light (Clocchiatti \etal\ 1995b).
Figure~\ref{fi:83n93j} shows that the V light curve of SN~Ib 1983N
was very similar to SN~1993J and very different than the SN~Ib 1994L.
Figure~\ref{fi:83v93j} shows that SN~1983V was a slow SN~Ic event and again
very similar to SN 1993J for perhaps 250 days.
We can see in Figure~\ref{fi:93jyics} that the light curves of SN~1993J 
were very similar to
the template light curves of slow SN~Ic for at least 110 days after
maximum, and some 250 days after maximum considering the B light curve of
SN~1983V.
This similarity of light curves and maximum brightness suggests that the slow
SN~Ic are an extreme version of the SN~Ib SN 1983N, which, in turn
may be a variation of the hybrid SN~1993J.
They are not members of a common spectroscopic class at maximum; 
however, a series of
spectroscopic links connecting them can be established {\em relying on the
similarity of light curves and maximum brightness as the assembling criteria}.
SN~1993J displayed strong H lines before the second maximum and strong He
lines starting approximately 15 days after the second maximum.
SN~1983N displayed strong He lines at maximum, and residual H$\alpha$.
The slow SN~Ic display residual He, as suggested by SN~1983V.
One is led to the suggestion that all of these SNe shared the same type of
inner structure, which would be responsible for the general appearance of the
light curves starting some 20 days after maximum, but that they
have different outer layers on top of that common inner region which give them
the different spectra near
maximum light and hence different spectroscopic categories.

This scenario receives support from theoretical studies of evolution of stars in
interacting binary systems.
Woosley \etal\ (1994) and Woosley, Langer, \& Weaver (1995) computed
presupernova evolution of helium stars that experience extensive
mass loss to a companion and explode by core collapse.
One of their results was that, for the range of parameters
of the binary that they considered, there is a convergence of the pre--collapse
mass and inner density structure of the progenitor.
For a large range of initial masses of the system the final mass of the
pre--collapse star falls in the interval 2.3--4.3 \m.
In particular, Woosley \etal\ (1994) provided models for SN~1993J.
Their preferred model, 13B, ejected approximately 2.3\m, of which
$\sim$1.5\m\ is He and $\sim$0.15\m\ is H.
Small variations of this model could
provide explosions with different spectra near maximum light, but still
maintain essentially the same light curve especially at intermediate and
late--times.
In this framework, SNe~1983N, 1983V, and 1993J could be the
observational evidence that the convergence of different initial conditions
towards the same kind of pre--supernova structure and explosion
actually exists. This cannot, however, be the whole story.

\section{Discussion} 
\label{se:cw:disc}

The suggestion that supernovae 
of hybrid type like SN~1993J, SN~Ib like
SN~1983N, and slow SN~Ic, have the same kind of inner
structure -- as revealed by their light curves -- but 
a sequence of smaller envelopes
-- as revealed by the decreasing importance of the H/He lines -- 
leads to a contradiction.
How can a progenitor like that of SN~1993J,
if it were a star like model 13B of Woosley \etal\ (1994), 
lose its small H envelope and most of its more substantial He envelope
to become a SN~Ic
and still produce a light curve which decays like SN~1993J for hundreds of
days?

\begin{figure}
\psfig{figure=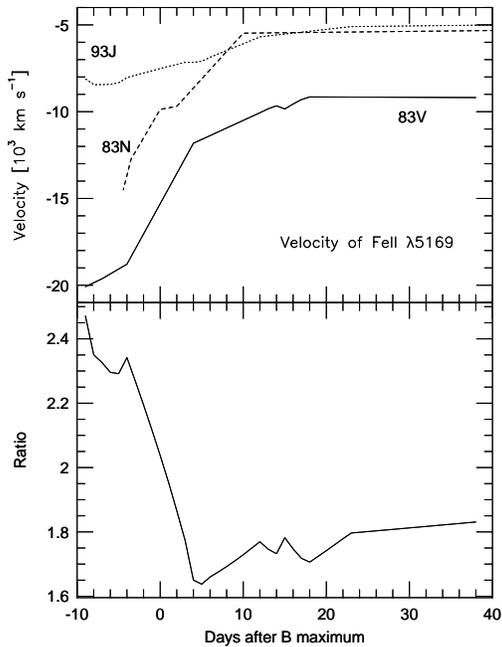,height=4in} 
\caption{\label{fi:83V93Jvel}  The evolution of the
photospheric velocity of
SN~1983V is compared to that of SN~1993J and SN~1983N as measured by
the line of FeII\l5169.}
\end{figure}

We can think of three possible answers to this question and two of them
contradict the observations of SN~1983V.
One is that the explosion mechanism is able to tune the amount of energy
transferred to the ejecta according to the mass of the ejecta (i.e. transfer a
constant amount of specific energy).
This possibility is discarded by the observations of the velocity in SN~1983V
(Clocchiatti \etal\ 1996c), which reveal that the mass to energy ratio was
larger in SN~1993J than in SN~1983V.  
As shown in Figure \ref{fi:83V93Jvel}, the photospheric velocity of
SN~1993J is always lower than that of SN~1983V which starts very high
but reaches a plateau parallel to that of SN~1993J about 10 days after
maximum.  The velocity ratio for the next 30 days is about a factor of 1.8.
For the same energy, this means that the ejected mass in SN~1983V was only
about one-third that of SN~1993J.  Thus the nearly identical light curves
are remarkable, indeed.

%%%%%%  Figure 7 here %%%%%%%%%%
%%%%%%%% Figure showing velocities of 93J 83V %%%%%%%%%%%% 

Another possible answer is that model 13B is wrong.
If SN~1993J were actually a massive C$+$O star with just a small amount of both
H and He in its external layers, then it might not be difficult 
to go from a hybrid SN~II to a SN~Ib or even to a SN~Ic 
without significatively altering the
light curve.  
A model for SN~1993J with some of these properties has
been proposed by Utrobin (1996).
Again, this possibility would again 
lead to explosions with nearly the same mass
to energy ratio, in contradiction with the velocities of SN~1983V.

The remaining possibility, is that
the late--time exponential decay does not depend on the mass to energy ratio but
is dominated by the $\gamma$--ray opacity of portions of the ejecta which do
not evolve following the expected homologous spherical expansion.
These portions of the inner ejecta would have to dominate 
the $\gamma$--ray opacity and  
the energy deposition starting some 40 days after maximum light. 
Their evolution in time would have to be independent of the details of the
structure of the outer layers.

It is suggestive in this regard that the oxygen--rich SN remnants 
do show structures in the high velocity oxygen--rich material.
On the basis of chemical abundances and
association with HII regions these remnants
have been proposed to result from SN~Ib
explosions (van den Bergh 1988, see also Weiler \& Sramek 1988).
The so--called oxygen knots most plausibly originated in the core of the
exploding star, have conserved their spatial identity
since explosion, and have 
expanded much less than the rest of the ejecta.
If a substantial fraction of the $^{56}$Ni generated in the explosion
remained trapped within structures of this kind, then the efficiency of the
ejecta to trap $\gamma$--rays would be independent of total mass of the ejecta,
or the mass to energy ratio.
Analysis of the chemical abundances in some of the oxygen knots in Cas~A led
Chevalier \& Kirshner (1978) to the conclusion that Fe was overabundant
by factors of 3 to 4 with respect to cosmic values, although the uncertainties
were large.
Other oxygen--rich SN remnants also show spatial structures that have been
described as luminous arcs of bright knots
(van den Bergh 1988), and planar or warped expanding rings
(e.g., Tuohy \& Dopita 1983).  The Cygnus Loop may also show 
structure related to this phenomenon (Miyata: these proceedings).

Regarding the fast SN~Ic, the evidence we have presented 
suggests that the explosions in which they originate must be
different in a substantial manner from the explosions that yield their
slow counterparts.
They may be similar to the slow SN~Ic, but have none of the hypothesized
non-homologous component so that their decay is strictly due to
the decrease in deposition of $\gamma$-rays in the low-mass expanding
ejecta.

The similarity of the spectral evolution and the dispersion in 
light curve properties of fast and slow SN~Ic 
suggests that we are confronting progenitor stars of a
similar class that have a substantial difference in the inner structure, or that
explode through a different explosion mechanism.
The fact that the absolute brightness at maximum is similar
suggests that the masses of $^{56}$Ni ejected in the
explosions are similar for both types and hence
that the explosion mechanism, presumably core collapse,
is essentially the same.

Some part of the inner
structure of the progenitors must then be significantly different in slow and
fast SN~Ic in order to make such a different light curve.
If it is true that dynamical instabilities in the inner ejecta of slow
SN~Ic play the fundamental role we have suggested above, then it is
possible that the reason for the different light curves is that fast
SN~Ic do not provide the appropriate background for these instabilities
to grow.
In such a case, their ejecta would follow much more closely the expected
decrease of the $\gamma$--ray opacity, proportional to the inverse square of
time, and their light curve will be faster.

The challenge would then be to identify the critical element that
will promote or quench the growth
of instabilities in the innermost portions of the ejecta of
a C$+$O star with little helium in the external layers that
explodes by core--collapse. 
It is important to emphasize that 
the light curves seem to cluster in two
classes with no intermediate slopes indicative of a continuum of cases.
On the other hand, whatever processes produce slow light curves, 
they seem to lead to a
remarkably homogeneous evolution of the $\gamma$--ray opacity at medium to
late--times.
This is an indication that the physical process that dominates the
$\gamma$--ray optical depth is fairly independent of those details of the
progenitor structure from events like SN~1993J to
those like SN~1983V.

We may finally try to consider how events like SN~1984L and SN~1985F fit
within this framework.
The maximum light spectrum of SN~1984L was very similar to that of SN~1983N
with no significant difference in the expansion velocities (see comparison
of maximum light spectra in Wheeler \etal\ (1994).
The evolution of the light curves around maximum was also similar,
but soon after maximum SN~1984L settled on an exponential
decay with a slope of 0.01~mag~day$^{-1}$, the minimum rate of decay for
supernovae powered by the decay of $^{56}$Co, and consistent with full trapping
of the $\gamma$~--rays generated in this decay.
This rate continued for more than 500 days
after maximum (Schlegel \& Kirshner 1989).
The spectrum of SN~1985F near maximum light is not known, but its B light curve
was similar to that of SN~1984L, especially in the slow late--time decay.
The usual interpretation of these light curves in terms of a large ejecta
mass covering the radioactive $^{56}$Co, but expanding at the rate
predicted by homologous expansion, led Swartz \& Wheeler (1991) to the
conclusion that the ejecta mass of SN~1984L was in excess of 10\m.

Could it be that instead of very massive ejecta, the explosion of these
objects produced a significant fraction of the ejecta which expanded very
little as the supernovae evolved?
SN~1985F was remarkable for the clear evidence of fine structure it showed
in its nebular lines, including the [OI] \l\l6300, 6364 doublet
(Filippenko \& Sargent 1989).  In addition,
its late--time spectrum indicates
that a macroscopic mixing of blobs from shells
with different nuclear products is required to match both
the relative fluxes and line profiles of the 
nebular lines (Fransson \& Chevalier 1989).
SN~1993J, however, with a light curve that decayed for the first hundreds of
days at a faster rate than those of SNe~1984L and 1985F, 
also displayed fine structure in the nebular lines.

Irrespective of the final answer to the problems posed by the light curves of
slow SN~Ic, they seem to indicate that spherically symmetric homologous
expansion does not apply to all the regions of the ejecta of core--collapse
supernovae, and that the late--time light curves give information on 
the physics of core collapse.

\acknowledgements 
JCW is grateful for the support of the meeting organizers to attend
this intensely interesting Advanced Study Institute.  We are
both grateful for the provision of data and for discussion 
of these topics with Stefano Benetti, Mark Phillips, Massimo Turrato,
and Stefano Cristiani.
This research is supported in part by NSF Grant AST 9218035 and NASA
Grants NAGW-2905 and NAG5-2888.
Support for AC was provided by
the National Science Foundation through grant GF-1001-95 from AURA, Inc.,
under NSF cooperative agreement AST-8947990, and from Fundaci\'on
Antorchas Argentina under project A-13313.

\end{document}